\documentclass[conference]{IEEEtran}
\ifCLASSINFOpdf
\else
\fi
\usepackage{graphics}
\usepackage[dvips]{epsfig}
\usepackage{boxedminipage}
\usepackage{rotating}
\usepackage{amssymb,amsmath,amsthm}
\newtheorem{theorem}{Theorem}

\newtheorem{lemma}{Lemma}

\newcommand{\cM}{{\cal M}}
\newcommand{\bC}{{\bf C}}

\begin{document}
%
\title{Anti-Structure Problems}

\author{
\IEEEauthorblockN{Ram Zamir}
\IEEEauthorblockA{
Department  of Electrical Engineering-Systems \\
Tel Aviv University, Israel \\
zamir@eng.tau.ac.il}
}


%


\maketitle


\begin{abstract}
The recent success of structured solutions for a class of
information-theoretic network problems, calls for exploring their
limits. We show that sum-product channels resist a solution by
structured (as well as random) codes.  We conclude that the
structured approach fails whenever the channel
operations do not commute
(or for general functional channels, when the channel function
is non decomposable).
\end{abstract}


\vspace{5mm}


\vspace{5mm}




%
\IEEEpeerreviewmaketitle


\section{Introduction}
Structured codes can be effective, and significantly better than
random codes, for various multiuser problems. Prominent examples
include the well known Korner-Marton (KM) ``modulo-two sum'' problem
\cite{KornerMarton79}, as well as more recent setups such as the
``dirty'' multiple-access channel (MAC)
\cite{Philosof2011},
the noisy linear network
(along with the compute \& forward relaying technique)
\cite{NazerGastparIEEEPROC},
and more \cite{ZamirPlenary2010}.
The effectiveness of structured codes, in
particular linear or lattice codes, is due to a good match between
their linear structure and the additive nature of the source or
channel involved.

In the Korner-Marton problem, for example,
the two components
$(X,Y)$ of a doubly-symmetric binary source are encoded separately.
The joint decoder is not interested in a full reconstruction of $X$
and $Y$, which are viewed as ``helper sources'',
but rather in their
modulo-two sum $X+Y$ (or Xor). Writing the statistical relation
between $X$ and $Y$ as a modulo-additive noise channel
\begin{equation}
\label{KM}
   Y = X + Z
\end{equation}
we can recast the problem as that of lossless reconstruction of the
noise $Z$ from separate encodings of $X$ and $Y$.

As shown by Korner and Marton,
a linear structured coding scheme,
which sends the syndromes of
$X^n$ and $Y^n$ with respect to a suitable
linear binary code, achieves the minimum possible rate of $H(Z)$
- the entropy of $Z$ - per each encoder.
In contrast, a conventional
{\em random} coding scheme cannot do better than to encode at a sum rate
equal to the joint entropy of $X$ and $Y$. This corresponds to the
Slepian-Wolf solution \cite{SlepianWolf73}, i.e., to a lossless reconstruction of
both $X$ and $Y$ at the decoder.
The resulting sum rate can be therefore
significantly higher than $2 H(Z)$ for
highly correlated sources.

The binary KM problem can be generalized to
a $q$-ary field,
in which case a linear $q$-ary code replaces
the linear binary code in
the KM solution \cite{HanKobayashi87}.
And it also has a quadratic-Gaussian version
\cite{PradhanKornerMarton}.


A dual example with a similar characteristics is that of the
``doubly dirty'' MAC.
This channel extends Costa's ``writing on a
dirty paper'' problem
\cite{Costa83,GelfandPinsker80}
to a MAC;
i.e., an additive-noise channel with two
inputs $X_1$ and $X_2$ and an output
$Y$ given by
\begin{equation}
\label{DMAC}
  Y = X_1 + X_2 + S_1 + S_2 + \mbox{noise}
\end{equation}
where $S_1$ and $S_2$ are two interferences, each known as ``side
information'' to one of the encoders. Addition in
(\ref{DMAC}) is over some
group in the discrete channel case, or the usual addition in the
continuous case. The problem is made interesting by imposing an
input constraint upon $X_1$ and $X_2$,
thus the encoders cannot simply
subtract the interferences.

Similarly to the Korner-Marton problem, a linear/lattice pre-coding
scheme (which subtracts the interference ``modulo the code'')
achieves the capacity region of this channel
\cite{PhilosofZamir2009,Philosof2011}.\footnote
{
In
the discrete noiseless case
the linear coding scheme is exactly optimal, while in
the continuous case it is asymptotically optimal
in the continuous high SNR case.
}
And in contrast, the rates achieved by a more
conventional random binning scheme {\em vanish} in the
limit of strong
interference signals.

This sharp discrepancy is due to the distributive
nature of the side-information;
if the knowledge of $S_1$ and $S_2$ were
{\em centralized} -
i.e., they were both known to one encoder or to the joint decoder,
then random binning could be effective and (nearly) achieve
capacity; see \cite{Philosof2011}.


Sometimes structured codes are {\em inferior} to random codes.  This
situation occurs when the linear structure of
the code causes
ambiguity at the decoder;
for example, the symmetric-rates point of
the (clean) MAC capacity region,
or of the Slepian-Wolf rate region.

In this short note we focus on another, perhaps obvious
weakness of structured
codes: they are sensitive to the structure of the channel.
Specifically, we show that if the additive channel in
(\ref{KM}) or in (\ref{DMAC}) is
replaced by a channel involving both {\em addition and
multiplication}, then structured codes - and in fact, any other
coding scheme - are not effective.


\section{Sum-Product Korner-Marton}
%
%
Consider a generalization of the KM problem (1),
where the statistical relation between the component
sources is given by the channel
\begin{equation}
\label{SumProductKM}
    Y = A + B \times C
\end{equation}
where $A,B$ and $C$ are statistically independent,
and $B \not= 0$ with probability one.
All variables in (\ref{SumProductKM}) belong to a finite
field $F_q$ of size $q$,
and the $+$ and $\times$ are the sum and product
operations over $F_q$.
Here $A$ and $Y$ are viewed as the ``helper sources'',
$C$ as the desired source,
and B as the ``channel state''.
The classical KM problem (\ref{KM}) thus corresponds
to the case where $q=2$,
$A = X$ is uniform over $\{0,1\}$,
$B = 1$, and $C = Z$.

In the {\em centralized state} case,
the channel state B is either known
to the joint decoder
(i.e., one encoder observes $A$, the other encoder
observes $Y$, and the decoder has access
to $B$),
or to both encoders
(i.e., one encoder observes $(A,B)$, and the other encoder
observes $(Y,B)$).
It is not hard to show that in this case
the compression rate is $H(Z)$,
independent of the state distribution,
as in the classical KM problem.\footnote
{
Regarding the former case ($B$ available at the decoder),
note that a random parity-check matrix
$H$ is ``good'' with high probability
for the classical KM problem
(\ref{KM})
(i.e., $Z^n$ can be reliably decoded from
the syndromes $H X^n$ and $H Y^n$);
hence, $H B^n$
is ``good'' with high probability
for the generalized KM problem
(\ref{SumProductKM}).
In the latter case,
the encoders simply divide their observations
by $B$,
hence get back to the classical KM problem.
}

Our focus is, however,
on the {\em de-centralized state} case,
where the channel state B is available
only to {\em one} encoder.
The performance in this case can be bounded
by the simplified setup shown in Fig. 1,
where one of the helper
sources is available (un-coded) as ``side information''
at the decoder.
That is,
there is only one encoder which observes
$X = (A,B)$,
while $Y$ is available at the decoder,
who wishes to reconstruct $Z = C$.

\begin{figure}
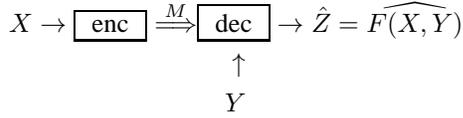

\begin{center}
\begin{align*}
X \rightarrow
\framebox{ enc } \stackrel{M}{\Longrightarrow}
& \framebox{ dec } \rightarrow \hat{Z}
= \widehat{F(X,Y)}
\\
& \ \ \ \uparrow
\\
& \ \ \ Y
\end{align*}
\caption{Functional source coding with side information
at the decoder.}
\label{fig:functionalSW}
\end{center}
\end{figure}

This latter problem is a special case of
{\em functional source coding}
\cite{HanKobayashi87},
where a function  $Z = F(X,Y)$
needs to be reconstructed from separate
coded versions of $X$ and $Y$.
%
The setup of Fig.~\ref{fig:functionalSW}
corresponds to the case
where $Y$ is given un-coded as ``side information''
to the decoder (or it is encoded at a rate
greater than or equal to $H(Y)$),
and where
\begin{eqnarray}
\nonumber
X = (A,B), \ \
Y = A + B \times C, \ \
Z = C
\\
\label{FunctionalWZ}
\mbox{and} \ \
F(X,Y) = (Y-A)/B.
\end{eqnarray}


%


A precise definition of functional
source coding with side information
at the decoder is as follows.
The encoding function is
$f: F_q^n  \rightarrow \cM$,
where the size of the message space $\cM$
is $2^{nR}$,
with $n$ being the code block length
and $R$ being the coding rate. 
The decoding function is
$g:  {\cal M} \times F_q^n \rightarrow F_q^n$.
The probability of error $P_e$ is
the probability that
$g(M, Y^n)$ is not equal to the vector $Z^n$,
where $Z_i = F(X_i,Y_i)$, $i=1 \ldots n$,
and where
$M = f(X^n)$
is the encoded message.
For a given memoryless double source
$(X_1,Y_1), (X_2,Y_2), ...$
and a function $F(.,.)$,
a rate $R$ is said to be ``achievable'' if
we can make $P_e$ as small as desired
for some functions $f$ and $g$,
and large enough $n$.
Finally, $R^*$ denotes the minimum
achievable rate.


Clearly, the minimum achievable rate R*
satisfies
\begin{equation}
\label{FunctionalWZbounds}
   H(X|Y) \geq R^* \geq H(F(X,Y)|Y)
\end{equation}
where the LHS corresponds to the case
where the decoder fully reconstructs $X$ before
computing $F(X,Y)$,
while the RHS corresponds to the case
where the encoder also has access to $Y$, so
it can first compute $F(X,Y)$ and then compress it.
In the sum-product case (\ref{FunctionalWZ}),
if $A$ is uniform over $F_q$,
then the bounds (\ref{FunctionalWZbounds})
become
%
\begin{equation}
\label{FunctionalWZboundsSumProduct}
   H(B) + H(C) \geq  R^*  \geq  H(C) .
\end{equation}
Note that in the classical KM problem
$B=1$, i.e, $H(B)=0$;
thus the bounds coincide,
and the coding rate is merely
the entropy of the desired variable $C$.


Han and Kobayashi \cite{HanKobayashi87}
give necessary and sufficient
conditions for the LHS of (\ref{FunctionalWZbounds}) to be tight.\footnote
{
They in fact consider a more general case
where both $X$ and $Y$ are encoded.
}
These conditions are satisfied in the sum-product case.


\begin{lemma}
\label{lemmaKM}
In the sum-product
(functional source coding)
problem (\ref{FunctionalWZ}),
$R^* = H(X|Y)$.
Thus, if $A$ is uniform over $F_q$,
then  $R^* = H(B) + H(C)$.
\end{lemma}

\begin{IEEEproof}
Follows since two different lines in $F_q$
intersect in at most one point,
implying the condition in
\cite[lem.1]{HanKobayashi87}.
\end{IEEEproof}


This result implies that the minimum coding rate
$R^*$ is in general larger than the entropy of the
desired variable $C$,
which is the rate in the classical KM setting (\ref{KM}).
In fact, the ``extra'' rate
can be as large as $\log (q-1)$,
for B which is uniform over $F_q \setminus 0$.

As a corollary from Lemma~\ref{lemmaKM},
it follows that the rate of the first encoder
in the distributed coding setup of (\ref{SumProductKM})
is at least $H(B) + H(C)$.
The interpretation is that
the introduction of
the multiplicative
state variable B breaks the symmetry
of the classical KM problem;
B must be fully conveyed to the decoder
before the linear structure of the channel
can be utilized
(by means of a linear ``syndrome'' coding)
to encode the desired source $C$.



\section{The Sum-Product Dirty MAC}
%
%
Consider next a modification of the
dirty MAC problem (\ref{DMAC}),
in which the channel output is given by
\begin{equation}
\label{DMACSumProduct}
    Y = A + A' + B \times C
\end{equation}
where as in (\ref{SumProductKM})
all variables belong to a finite
field $F_q$ of size $q$,
and the $+$ and $\times$ are the sum and product
operations over $F_q$.
The inputs of this MAC are $A$ and $A'$
(corresponding to $X_1$ and $X_2$ in (\ref{DMAC})),
while $B$ and $C$ are the channel state variables
(corresponding to $S_1$ and $S_2$ in (\ref{DMAC})).
There is no additional noise,
nor input constraints.

As in the sum-product KM,
the centralized state case is easy:
if both state variables $B$ and $C$ are known
to one encoder, or are known to the joint decoder,
then the product $B \times C$ can be simply
subtracted;
hence the sum capacity is $\log(q)$,
as if the channel was noiseless.

The interesting setup is, again, the
de-centralized state case.
That is,
each encoder has access to only one of the channel
states,
while the decoder is completely ignorant of the states.
The capacity in this case is bounded
from above by that of the {\em single-user} channel
shown in Fig. 2, with
\begin{equation}
\label{DMACSumProductSimple}
   Y = A + B \times C  .
\end{equation}
Here
there is a single encoder that has access to
one of the states ($S_1 = B$), while the decoder
has access to the second state ($S_2 = C$),
where $C$ is independent of both the
input $X=A$ and $B$.\footnote
{
A continuous version of this setup may be
thought of as a channel with a fading interference
\cite{AvnerZaidelShamaiErez}.
}



\begin{figure}
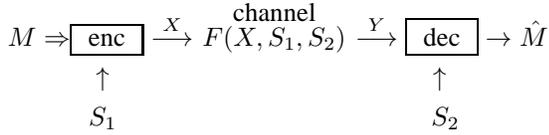

\begin{center}
\begin{align*}
M \Rightarrow
& \framebox{ enc } \stackrel{X}{\longrightarrow}
\
\stackrel{ \mbox{channel} }{ F(X,S_1,S_2) }
\
\stackrel{Y}{\longrightarrow}
\framebox{ dec } \rightarrow \hat{M}
\\
&  \ \ \uparrow
\ \ \ \ \ \ \ \ \ \ \ \ \ \ \ \ \ \
\ \ \ \ \ \ \ \ \ \ \ \ \ \ \ \ \
\uparrow
\\
& \  \  S_1
\ \ \ \ \ \ \ \ \ \ \ \ \ \ \ \ \
\ \ \ \ \ \ \ \ \ \ \ \ \ \ \ \ \
S_2
\end{align*}
\caption{
A deterministic channel with two states,
one known to the encoder and another known
to the decoder.}
\label{fig:functionalGP}
\end{center}
\end{figure}
%


A precise definition of encoding
and decoding over such a channel
is as follows.
The encoding function is
$f: M  \times  F_q^n  \rightarrow F_q^n$
and
the decoding function is
$g: F_q^n  \times  F_q^n  \rightarrow \cM$,
where the size of the message space
$\cM$ is $2^{n R}$ , $R$ being the coding rate.
The error probability $P_e$ is the
probability that
$g(Y^n, S_2^n) \not= M$,
where
$Y^n$ depends on $X^n$, $S_1^n$ and $S_2^n$,
and where
$X^n = f(M, S_1^n)$
for $M \in \cM$.
A rate $R$ is said to be ``achievable'' if
we can make $P_e$ as small as desired
for some functions $f$ and $g$,
and large enough $n$.
Finally,
the capacity $\bC$ is the highest achievable
rate.


The sum-product channel (\ref{DMACSumProductSimple}) is,
in fact,
a deterministic channel,
where the output $Y$ is a function
of the input $X$, and the two states
$S_1$ and $S_2$:
%
\begin{equation}
\label{FunctionalGP}
Y = F(X,S_1,S_2).
\end{equation}
There is no additional noise in the channel,
beyond the randomness of the two (known) states $S_1$ and $S_2$.

The setup of (\ref{FunctionalGP}) is an instance
of the Gelfand-Pinsker problem
\cite{GelfandPinsker80},
i.e., a channel with non-causal side information
at the encoder.
Hence, it has a single letter solution of the
form
\begin{equation}
\label{GP}
\bC = \max { I(U; Y, S_2) - I(U; S_1) }
\end{equation}
where the maximization is over a suitable
set of auxiliary random variables $U$,
and functions  $X = X(U,S_1)$.\footnote
{
The admissible $U$'s are those for which
$S_2$ is independent of  $(U,X,S_1)$,
and
$U \leftrightarrow (X,S_1) \leftrightarrow Y$
form a Markov chain for each value of $S_2$.
Since $S_2$ is independent of $U$,
the first term in (\ref{GP}) can be written
also as  $I(U; Y | S_2)$ .
}


The structure of the function $F$
in (\ref{FunctionalGP}) plays a key role in determining
the capacity $\bC$.
A favorable case is when $F$ has a {\em composite
form}, where the dependence on the
encoder variables $(X,S_1)$ is separate
from the decoder state $S_2$.


\begin{lemma}
If the function $F$ can be decomposed into
$F(a,b,c) = \tilde{F}(G(a,b),c)$,
where
$\tilde{F}$ is invertible with respect to the
first argument
(i.e., the equation $y = \tilde{F}(t,c)$ has a solution
$t$ for every $y$ and $c$),
then (\ref{GP}) is optimized by $U = G(X,S_1)$.
If also $G$ is invertible with
respect to the first argument,
then the capacity is
\[
\bC = \log(q)
\]
and it is achieved by an input $p(x|s_1)$
that makes $G(X, s_1)$ uniform
for all values of $s_1$.
\end{lemma}

\begin{IEEEproof}
The first part
follows from \cite[sec.~III.F]{BCW03},
and the invertibility of $\tilde{F}$.
See also \cite{EliHaimMS}.
\end{IEEEproof}


The sum-product channel (\ref{DMACSumProductSimple})
clearly does not satisfy the first condition
of the lemma, as addition and multiplication
do {\em not} commute.
In fact, this channel is much worse.
To assess its capacity,
we shall first establish a relation
to a ``minimum entropy'' problem.


\begin{lemma}
The capacity of a deterministic two-state channel
of the form
\begin{equation}
\label{FunctionalGPadditive}
   Y = X + F(S_1,S_2)
\end{equation}
where $S_1$ and $S_2$ are available at the encoder
and the decoder, respectively,
is given by
\begin{equation}
\label{MinEntropy}
\bC = \log(q) - \inf \frac{1}{n}
H \Bigl( g(S_1^n ) + F(S_1^n, S_2^n) | S_2^n \Bigr)
\end{equation}
where
the second term is the (average) conditional entropy
given $S_2^n$,
and where
the infimum is over all code block lengths $n$,
and functions
$g: F_q^n  \rightarrow F_q^n$.
\end{lemma}

\begin{IEEEproof}
Easy and will be omitted.
\end{IEEEproof}


Although (\ref{MinEntropy}) is not a single-letter expression,
it is sometimes easier for analysis than
the Gelfand-Pinsker solution (\ref{GP}).
Specifically,
for the sum-product channel (\ref{DMACSumProductSimple}) we have:


\begin{theorem}
{\bf (Shany-Zamir \cite{ShanyZamir})}
For the case where
$Y = X + S_1 \times S_2$,
the minimum entropy in (\ref{MinEntropy})
is bounded by  
\[
\log \left( \frac{q}{2} \right)
\leq
H_{min}
\leq
\log \left( \frac{q}{2-1/q} \right)
\]
where
the upper bound is achieved
(for all $n$)
by a quadratic per-letter
function
$g(s) = s^2$.
\end{theorem}


As a corollary from this theorem,
we conclude that the capacity of the sum-product
channel (\ref{DMACSumProductSimple})
is {\em at most one bit}, for all $q$.
This is quite disappointing when compared to the
capacity of $\log(q)$, which is achievable in the
centralized-state case.
The same statement is true also for the
sum-product dirty MAC (\ref{DMACSumProduct}).

%
\section{Conclusions}
%
%
The essence of the examples given in this paper is that
the order of performing the sum and product operations
matters;
and in fact, they are very ``different''.
One aspect of this difference -
which is related to the sum-product KM problem -
is that the expression
\begin{equation}
\label{sum_product}
  a + b \times c
\end{equation}
cannot be decomposed into the form:
$\mbox{function( function $(a,b),c)$}$
(not even approximately).
A second aspect -
related to the sum-product dirty MAC -
is that it is impossible to find
a function of $a = a(b)$ such that (\ref{sum_product})
would be only a function of $c$
(not even approximately).
In contrast,
these two requirements are easily fulfilled
if the expression in (\ref{sum_product}) is a pure sum
  $a + b + c$
(by the associativity of summation),
or a pure product
  $a \times b \times c$
(by the associativity of multiplication).\footnote
{
Note that pure multiplicative versions of the
generalized KM and DMAC problems
can be solved using linear codes over
a ``logarithmic'' domain.
}

It would be interesting to explore further
(and perhaps quantify)
the information-theoretic aspects
of function decomposition.
Note that this question is
almost ``distribution free''
(i.e., nearly independent of the
probability distributions of sources
and channels).
A different aspect of ``anti structure'',
which is due to a ``bad'' noise distribution,
can be found in \cite{CohenZamir}.





\section*{Acknowledgment}
I thank Yaron Shany for many interesting discussions,
and to Prakash Iswar and Sandeep Pradhan for pointing
out the relation to the work of Han and Kobayashi.


\vspace{5mm}

\bibliographystyle{IEEEtran}




%

\end{document}